\documentclass{article}

\usepackage{amsmath}
\usepackage[utf8]{inputenc}
\usepackage[T1]{fontenc}
\usepackage{geometry}
\usepackage{enumerate}
\title{Quantum Particle Dynamics in a Highly Singular 1D-Potential $U(x) = -\alpha \delta(x) + \beta \delta'(x)$ Superposed on a Well-Behaved One}
\author{ }
\date{ }
\begin{document}
\maketitle
\vspace{-30pt}
\begin{flushleft}
\center Norman J. Morgenstern Horing*\\
\center Department of Physics and Engineering Physics\\
\center Stevens Institute of Technology\\
\center Hoboken, NJ 07030, USA\\
\center March 1, 2013
\end{flushleft}
\begin{abstract}
We examine the one-dimensional quantum dynamics of a Schr$\ddot{o}$dinger particle in a potential represented by a generalized  function of the form $U(x) = -\alpha \delta (x) + \beta d(\delta (x))/dx$ superposed on a well behaved potential $V(x)$. In this, we construct the full, exact Green's function for such a 1D system analytically in closed form, taking account of a spatially variable mass $m(x)$. Our result shows that there can be no electron transmissions through the $\beta \delta '(x)$- potential, regardless of the presence of the $V(x)$- potential and $\alpha \delta (x)$, (with $\alpha \ne 0$).
\end{abstract}
\section{Introduction$^{1-5}$}
\begin{paragraph}
\ The advent and rapid development of the fabrication of low dimensional semiconductor materials, replete with the promise of nanostructures upon which a whole new generation of quantum electronic and optical devices can be based, has stimulated an enormous effort to explore the physical properties of such materials, and how they can be manipulated to greatest advantage. Practically all the fields of science and engineering are involved in this massive effort throughout the world. Mathematical modelling has an important role in this matter, enabling analyses that provide insight into the quantum mechanical behavior of nanostructures and their possible optimization. One avenue of such studies over the past quarter century has been the introduction of generalized functions into the potential involved in nonrelativistic one-dimensional Schr$\ddot{o}$dinger dynamics, in particular, the Dirac delta function and its derivative. The inclusion of the Dirac delta function $(\delta (x))$ itself as a potential has proven to be quite straightforward; however,  the inclusion of its derivative $(\delta '(x))$ has brought forth substantial controversy. As far back as 1986, Seba$^4$ found that electron transmission through that highly singular potential could not take place. However, other researchers have appended boundary conditions to $\delta '(x)$ and claimed that electron transmission can occur.
\end{paragraph}
\begin{align*}
^{*} &\ \mbox{Corresponding Author: N.J.M Horing, phone 201-216-5651; fax 201-216-5638; email} \\ &\mbox{<nhoring@stevens.edu>.}
\end{align*}
\begin{paragraph}
\ We recently studied this problem by constructing the full, exact Green's function for the Dirac-delta-function-derivative model, defining it solely in terms of the usual derivative property [$\delta '(x) \equiv d(\delta(x))/dx$] under integration by parts, with no appeal to additional boundary conditions.  Aside from such an integration by parts, we employ only the conventional, well established properties of the Dirac delta function, $\delta(x)$, in terms of integration (as a generalized function), $\int_{-\infty}^{\infty}dx \: \delta (x-a) f(x) = f(a)$, and differentiation, $ d(\eta_{+}(x-a))/dx = \delta(x-a)$,where $\eta_{+}(x)$ is the Heaviside unit step function (which is understood to have the value $\eta_{+}(0) = 1/2$, to which a Fourier series representation converges at the position of the step).  Our avoidance of boundary conditions is due to the fact that they are not at all necessary under the usual conventions and that they may distort the meaning of $\delta^{\prime}(x)$ beyond recognition.  The Green's function we obtain analytically in closed form confirms Seba's finding that there can be no electron transmission through the $\delta '(x)$- singular potential if the potential profile is otherwise well-behaved, and that such a wave-packet must be totally reflected.

\indent In the present paper we extend these one-dimensional considerations to take into account a spatially variable mass, $m(x)$, and a reasonably well behaved spatially variable potential, $V(x)$, and a $\delta(x)$- potential, all in addition to the $\delta '(x)$- potential previously examined. It is in this very general situation that we construct the full, exact Schr$\ddot{o}$dinger Green's function in closed form, and show that \emph{no} electron wave packet transmission through $\delta '(x)$ can occur.  Our analysis of this problem is in agreement with aspects of the formulation by Park$^{14}$, but the later is quite limited in application to one specific case.
\end{paragraph}
\section{Derivation of the Exact One-Dimensional Green's Function}

\indent Allowing for a variable mass, $m(x)$, the one-dimensional Schr$\ddot{o}$dinger Green's function equation for a system with time-translational invariance has the Sturm-Liouville form in frequency representation $t-t' \rightarrow \omega$ and we suppress the explicit appearance of $\omega$, so $G(x,x'; t-t') \rightarrow G(x,x';\omega) \equiv G(x,x')$ as given by$^{16}$
\begin{equation}
  \left[\frac{\partial}{\partial x} \left( \frac{1}{m(x)} \frac{\partial}{\partial x} \right) + V(x) + U(x) \right]G(x,x') = \delta (x-x').\tag{1}
\end{equation}
Here, $V(x)$ is understood to be a relatively well behaved potential (which can accommodate finite discontinuities using the well known Green's function joining technique employed in the theory of surface/interface states$^{17-21}$; it also includes an $\omega$ - term from the Fourier time-transform) and all highly singular delta-function-type potentials are relegated to $U(x)$ as
\begin{equation}
  U(x) = -\alpha \delta(x) + \beta \delta'(x)\tag{2}
\end{equation}
\begin{center}
($\alpha$, $\beta$ are constants and $\delta'(x) \equiv d(\delta(x))/dx$).
\end{center}
\indent To start, we define an auxiliary Green's function, $G_{0}(x,x')$, as the inverse of the Sturm-Liouville operator excluding $U(x)$:
\begin{equation}
  \left[\frac{\partial}{\partial x} \left( \frac{1}{m(x)} \frac{\partial}{\partial x} \right) + V(x) \right]G_{0}(x,x') = \delta (x-x')\tag{3}
\end{equation}
Taking $y_{1}(x)$ and $y_{2}(x)$ as two linearly independent solutions of the homogeneous counterpart of Eq. (3), with $y_{1}$ chosen to satisfy the boundary condition at the lower limit and $y_{2}$ doing so at the upper limit, the solution of Eq.(3) is known to be
\begin{equation}
    G_{0}(x,x') = \frac{m(x')}{\Delta (y_{1},y_{2})} \left\{
    \begin{aligned}
      & y_{1}(x)y_{2} (x') \text{ for } x<x'\\
      & y_{2}(x)y_{1} (x') \text{ for } x>x'
    \end{aligned}
\right\},  \tag{4}
\end{equation}
where $\Delta(y_{1},y_{2})$ is the Wronskian of the two solutions, $y_{1}(x'),y_{2}(x')$, evaluated at $x'$:
\begin{equation}
  \Delta(y_{1}, y_{2}) = \text{det} \left|
  \begin{aligned}
  & y_{1} \text{  } y_{1}'\\
  & y_{2} \text{  } y_{2}'
  \end{aligned}
  \right| , \tag{5}
\end{equation}
($y' \equiv dy(x')/dx'$).\\
Considering $y_{1}$, $y_{2}$ to be known and hence $G_{0}$ is known, Eq. (1) may be rewritten as
\begin{equation}
  G(x,x') = G_{0}(x,x') + \int dx''G_{0}(x,x'')U(x'')G(x'',x'),\tag{6}
\end{equation}
or
\begin{equation}
  G(x,x') = G_{0}(x,x') - \alpha G_{0}(x,0)G(0,x') + \beta \int dx''\delta'(x'')G_{0}(x,x'')G(x'',x').\tag{7}
\end{equation}
Integrating by parts, this becomes
\begin{equation}
  G(x,x') = G_{0}(x,x') - \alpha G_{0}(x,0)G(0,x') - \beta \int dx''\delta(x'')\frac{\partial}{\partial x''}\left[G_{0}(x,x'')G(x'',x')\right]. \tag{8}
\end{equation}
Introducing the notation

\begin{align*}
\frac{\partial}{\partial x''}G(x'',x') &\equiv \left[\partial_{(L)} G(x'',x')\right]; \frac{\partial}{\partial x''} G(x',x'') \equiv \left[\partial_{(R)} G(x',x'')\right]; \notag \\
&\frac{\partial}{\partial x'} \frac{\partial}{\partial x''} G(x',x'') \equiv \left[\partial^2_{(L,R)} G(x',x'')\right]. \tag{9}
\end{align*}
Eq.(8) may be written as
\begin{align}
G(x,x') = G_0(x,x') - \alpha G_0 (x,0) G(0,x') - \beta \left[\partial_{(R)} G_0(x,0)\right] G(0,x') \notag \\ - \beta G_0(x,0) \left[\partial_{(L)} G(0,x')\right]. \tag{10}
\end{align}
To solve, we need to determine $G(0,x')$ and $\left[\partial_{(L)} G(0,x')\right]$: Setting $x\to{0}$, we obtain Eq.(10) as
\begin{align}
G(0,x') = G_0(0,x') - \alpha G_0(0,0) G(0,x') - \beta \left[\partial_{(R)} G_0 (0,0)\right] G(0,x')\notag \\ - \beta G_0(0,0) \left[\partial_{(L)} G(0,x')\right]. \tag{11}
\end{align}
Forming the left derivative of Eq.(10), we have
\begin{align}
\left[\partial_{(L)} G(x,x')\right] &= \left[\partial_{(L)} G_0(x,x')\right] - \alpha \left[\partial_{(L)} G_0 (x,0) \right] G(0,x') \notag \\ &- \beta \left[\partial^2_{(L,R)} G_0(x,0) \right] G(0,x') - \beta \left[\partial_{(L)} G_0(x,0)\right] \left[\partial_{(L)} G(0,x')\right], \tag{12}
\end{align}
and putting $x\to{0}$ in Eq.(12), we have
\begin{align}
\left[\partial_{(L)} G(0,x')\right] &= \left[\partial_{(L)} G_0(0,x')\right] - \alpha \left[\partial_{(L)} G_0(0,0)\right] G(0,x') \notag \\ &- \beta \left[\partial^2_{(L,R)} G_0 (0,0) \right] G(0,x')
 - \beta \left[\partial_{(L)} G_0(0,0)\right] \left[\partial_{(L)} G(0,x')\right]. \tag{13}
\end{align}
which expresses $\left[\partial_{(L)} G(0,x')\right]$ in terms of $G(0,x')$ as
\begin{align*}
\bigg(1+ \beta \left[\partial_{(L)} G_0(0,0)\right]\bigg) \left[\partial_{(L)} G(0,x')\right] = [\partial_{(L)} G_0(0,x')] - \alpha \left[\partial_{(L)} G_0(0,0)\right] G(0,x') \\ - \beta \left[\partial^2_{(L,R)} G_0(0,0)\right] G(0,x'), \tag{14}
\end{align*}
or
\begin{align}
&\left[\partial_{(L)} G(0,x')\right] = \bigg(1+\beta \left[\partial_{(L)} G_0(0,0)\right] \bigg)^{-1} \notag \\ &\times \bigg(\left[\partial_{(L)} G_0(0,x')\right] - \alpha \left[\partial_{(L)} G_0(0,0)\right] G(0,x')
- \beta \left[\partial^2_{(L,R)} G_0 (0,0)\right] G(0,x')\bigg). \tag{15}
\end{align}
Employing this result in Eq.(11), we have
\begin{align}
&G(0,x') = G_0(0,x') - \alpha G_0(0,0) G(0,x') - \beta \left[\partial_{(R)} G_0(0,0)\right] G(0,x')
 \notag \\& - \beta G_0(0,0) \bigg(1+ \beta \left[\partial_{(L)} G_0(0,0)\right]\bigg)^{-1} \bigg\{ \left[\partial_{(L)} G_0(0,x')\right] - \alpha \left[\partial_{(L)} G_0(0,0)\right] G(0,x') \notag \\& - \beta \left[\partial^2_{(L,R)} G_0(0,0)\right] G(0,x')\bigg\}, \tag{16}
\end{align}
which yields $G(0,x')$ as
\begin{align}
&G(0,x') = \bigg\{ 1+\alpha G_0(0,0) + \beta \left[\partial_{(R)} G_0 (0,0)\right] - \beta G_0(0,0) \bigg(1+\beta \left[\partial_{(L)} G_0(0,0)\right]\bigg)^{-1} \notag \\
&\times \bigg( \alpha \left[\partial_{(L)} G_0(0,0)\right] + \beta \left[\partial^2_{(L,R)} G_0(0,0)\right]\bigg)\bigg\}^{-1} \notag \\
&\times \bigg[G_0 (0,x') - \beta G_0 (0,0) \bigg(1+\beta\left[ \partial_{(L)} G_0(0,0)\right] \bigg)^{-1} \left[\partial_{(L)} G_0 (0,x')\right] \bigg]. \tag{17}
\end{align}
Eq. (17) may now be substituted into the right side of Eq. (15) to obtain $[\partial_{(L)}G(0,x')]$ in terms of $G_{0}$ and its derivatives alone. Finally, the substitution of these results for $G(0,x')$ and $[\partial_{(L)}G(0,x')]$ as indicated above into the right side of Eq. (11) yields the full, \emph{exact} Green's function for the highly singular 1-D potential $U(x)$ of Eq. (2) joined onto any relatively well behaved 1-D potential $V(x)$, such as a harmonic oscillator and/or electric field potential independent of time. For the special case of $\beta = 0$, we obtain
\begin{equation}
  G(x,x') = G_{0}(x,x') - \frac{\alpha G_{0}(x,0)G_{0}(0,x')}{1+\alpha G_{0}(0,0)}.\tag{18}
\end{equation}
\section{The Role of Highly Singular Potentials in the 1-D Green's Function with a well-behaved potential $V(x)$}
\indent To examine the role of the highly singular potentials in the 1-D Green's function, we rewrite Eq. (4) in the form
\begin{equation}
  G_{0}(x,x') = C(x') \{\eta _{+}(x'-x)y_{1}(x)y_{2}(x') + \eta _{+}(x-x')y_{2}(x)y_{1}(x')\} \tag{19}
\end{equation}
where $\eta_{+}(x)$ is the Heaviside unit step function and $C(x')\equiv m(x')/\Delta(y_{1},y_{2})$. Since $\eta_{+}(0) = 1/2$,
\begin{equation}
  G_{0}(0,0) = C(0)\left\{\frac{1}{2}y_{1}(0)y_{2}(0)+\frac{1}{2}y_{2}(0)y_{1}(0)\right\} = C(0)y_{1}(0)y_{2}(0).\tag{20}
\end{equation}
Differentiating Eq. (19) to form $[\partial_{(L)}G_{0}(x,x')]$, there is a cancellation of terms involving Dirac delta functions arising from $\partial_{(L)}\eta_{+}(x-x')$, etc., and we obtain
\begin{equation}
  [\partial_{(L)}G_{0}(x,x')] = C(x') \{\eta_{+}(x'-x)y'_{1}(x)y_{2}(x') + \eta_{+}(x-x')y'_{2}(x)y_{1}(x')\}. \tag{21}
\end{equation}
Differentiating again to form $[\partial^{2}_{(L,R)}G(x,x')]$, we have
\begin{align*}
  [\partial^{2}_{(L,R)}G_{0}(x,x')] &= C'(x')\{\eta_{+}(x'-x)y'_{1}(x)y_{2}(x') + \eta_{+}(x-x')y'_{2}(x)y_{1}(x')\}\\
  &+ C(x') \left\{
    \begin{aligned}
      &\delta(x'-x)y'_{1}(x)y_{2}(x') + \eta_{+}(x'-x)y'_{1}(x)y'_{2}(x') \notag \\
      &-\delta(x-x')y'_{2}(x)y_{1}(x') + \eta_{+}(x-x')y'_{2}(x)y'_{1}(x') \notag
    \end{aligned}
  \right\}, \tag{22}
\end{align*}
which may be written more compactly using the definition of the Wronskian, $\Delta_{x'}(y_{1},y_{2})$, as (subscript "$x'$" denotes evaluation at $x'$)
\begin{align*}
  [\partial^{2}_{(L,R)}G_{0}(x,x')] = &-C(x') \delta(x'-x) \Delta_{x'}(y_{1},y_{2}) \\
  &+ C(x')[\eta_{+}(x'-x)y'_{1}(x)y'_{2}(x') + \eta_{+}(x-x')y'_{2}(x)y'_{1}(x')] \\
  &+ C'(x')\{ \eta_{+}(x'-x)y'_{1}(x)y_{2}(x') + \eta_{+}(x-x')y'_{2}(x)y_{1}(x') \}.\tag{23}
\end{align*}
Clearly, the first term on the right of Eq. (23) shows that
\begin{equation}
  |[\partial^{2}_{(L,R)}G_{0}(0,0)]| \rightarrow \infty. \tag{24}
\end{equation}
\indent In view of the huge value of $|[\partial^{2}_{(L,R)}G_{0}(0,0)]|$, we may write Eq. (17) as
\begin{align*}
  G(0,x') = &- \frac{1+\beta[\partial_{(L)}G_{0}(0,0)]}{\beta^{2}G_{0}(0,0)[\partial^{2}_{(L,R)}G_{0}(0,0)]}
  \left\{ G_{0}(0,x') - \frac{\beta G_{0}(0,0)[\partial_{(L)}G_{0}(0,x')]}{1+\beta [\partial_{(L)}G_{0}(0,0)]}\right\}, \tag{25}
\end{align*}
and as long as we consider particle transmission through the highly singular potential at the origin (with $x<0$ and $x'>0$), this means that (Eq. (24))
\begin{align}
G(0,x') = 0. \tag{26}
\end{align}
On the same basis Eq. (15) for $[\partial_{(L)}G(0,x')]$ may be written as
\begin{equation}
  [\partial_{(L)}G(0,x')] = D^{-1}[\partial_{(L)}G_{0}(0,x')] + \frac{1}{\beta G_{0}(0,0)} \bigg\{ G_{0}(0,x')-\beta D^{-1}G_{0}(0,0) [\partial_{(L)}G_{0}(0,x')]\bigg\}, \tag{27}
\end{equation}
where we have defined $D$ as
\begin{equation}
  D = 1+\beta [\partial_{L}G_{0}(0,0)].\tag{28}
\end{equation}
It should be noted that the role of the singular potential part $\Delta U = -\alpha \delta (x)$ is in fact, negligibly small when $\beta \ne 0$, as indicated in our earlier work proving that there is no particle transmission across the $\beta \delta '(x)$- potential in the case of spatial translational invariance$^{22}$. Here, we see that $\Delta U = -\alpha \delta (x)$ remains negligible in $G(x,x')$ for well behaved potentials $V(x)$ so long as $\beta \ne 0$:
\begin{align}
G(x,x') = G_0 (x,x') &- \beta G_0(x,0) \bigg(D^{-1}[\partial_{(L)}G_0(0,x')] \notag \\ &+ \frac{1}{\beta G_0(0,0)}  \bigg\{G_0(0,x') - \beta D^{-1} G_0(0,0) [\partial_{(L)} G_0(0,x')] \bigg\}\bigg), \notag
\end{align}
or
\begin{align}
\hspace{5pt} G(x,x') = G_0(x,x') - \frac{G_0(x,0) G_0(0,x')}{G_0(0,0)} \tag{29}
\end{align}
\indent Considering that the Fourier transform of $G(x,x';\omega)$ to direct time representation represents the quantum mechanical amplitude for a Schr$\ddot{o}$dinger particle (wave packet) to be transmitted from position $x'$ at time $t'$ to $x$ at a later time $t$, such that
\begin{align}
\Psi_{out} (x,t) = \int \frac{d\omega }{\pi} e^{-i \omega(t-t')} \int^{\infty}_{-\infty} dx' G(x,x';\omega) \Psi_{in} (x',t'), \tag{30}
\end{align}
it is clear that an electron wave packet, $\Psi_{in} (x'<0,t' )$ in the region of incidence, $x'<0$, cannot be transmitted to $\Psi_{out} (x>0,t)$ in the outgoing wave region, $x>0$, on the other side of the highly singular potential $\beta \delta '(x)$ because Eq. (4) yields
\begin{align}
G(x>0, x'<0; \omega) &= C(x') \bigg[y_1 (x) y_2 (x') - \frac{y_1 (x) y_2(0) y_1(0) y_2(x')}{y_1(0)y_2(0)}\bigg] \notag \\
& \equiv 0. \tag{31}
\end{align}
This very general result means that there is \emph{no} possibility of particle transmission from $x'<0$ through the $\beta \delta '(x)$- potential, even in the presence of well behaved potentials $V(x)$ as well as the presence of $\Delta U = -\alpha \delta(x)$.
\newpage

\section{References}
\begin{enumerate}
  \item P. L. Christiansen, H. C. Arnbak, A. V. Zolotaryuk, V. N. Ermakov and Y. B. Gaididei; J. Phys. A: Math. Gen. \textbf{36}, 7589 (2003)\textbf{.}
  \item F. A. B. Coutinho, Y. Nogami, J. Fernando Perez; J Phys. A. Math. Gen. \textbf{30}, 3937 (1997)\textbf{.}
  \item Yuriy Golovaty; arXiv:1201.2610v1 [math.SP] 12 Jan 2012\textbf{.}
  \item Petr \u{S}eba; Reports on Math. Phys. \textbf{24}, 111 (1986)\textbf{.}
  \item F. M. Toyama and Y. Nogami; J. Phys. A: Math. Theor. \textbf{40}, F685 (2007)\textbf{.}
  \item M. Gadella, M. L. Glasser, L. M. Nieto; Int. J. Theor. Phys. \textbf{50}, 2144 (2011)\textbf{.}
  \item M. Gadella, J. Negro, L. M. Nieto; Physics Letters A \textbf{373}, 1310 (2009)\textbf{.}
  \item M. Gadella, M. L. Glasser, L. M. Nieto; Int. J. Theor. Phys. \textbf{50}, 2191 (2011)\textbf{.}
  \item J. J. Alvarez, M. Gadella, M. L. Glasser, L. P. Lara, L. M. Nieto; Journal of Physics: Conference Series \textbf{284}, 012009 (2011)\textbf{.}
  \item A. V. Zolotaryuk, Y. Zolotaryuk; arXiv:1202.1117v1 [math-ph] 6 Feb 2012\textbf{.}
  \item Johnannes F. Brasche, Leonid Nizhnik; arXiv:1112.2545v1 [math.FA] 12 Dec 2011\textbf{.}
  \item P. Kurasov; Math. Analysis and Applic. \textbf{201}, 297 (1996)\textbf{.}
  \item S.Kocinak and V.Milanovic, Modern Physics Letters B, \textbf{26}, 1250092 (2012).
  \item D. K. Park, J. Phys. A Math.: Gen. \textbf{29}, 6407 (1996).
  \item Haydar Uncu, Devrim Tarhan, Ersan Demiralp, \"{O}zg\"{u}r E. M\"{u}stecapho\u{g}lu; Phys. Rev. A \textbf{76}, 013618 (2007)\textbf{.}
\item P. M. Morse and H. Feschbach, "Methods of Theoretical Physics", Vol \textbf{1}, McGraw-Hill Book Company, Inc., 832 (1953).
\item F. Garcia-Moliner and F. Flores, "Introduction to the Theory of Solid Surfaces", Cambridge University Press (1979).
\item F. Garcia-Moliner and V. R. Velasco, "Theory of Single and Multiple Interfaces", World Scientific (1992).
\item S. G. Davison and J. D. Levine, "Solid State Physics", Vol. \textbf{25}, Eds: H. Ehrenreich, F. Seitz, and D. Turnbull, Academic Press (1970).
\item S. G. Davison and M. Steslicka, "Basic Theory of Surface States", Clarendon Press, Oxford (1992).
\item J. Inglesfield, J. Phys. C; Solid State Phys. 4, L 14, (1971).
\item N.J.M Horing, unpublished.
\end{enumerate}
\end{document}